\pdfoutput=1

\documentclass[aps,prl,preprint,superscriptaddress,amsfonts,amssymb,amsmath,showpacs,titlepage,a4paper]{revtex4-1}

\usepackage{latexsym}
\usepackage{verbatim} 
\usepackage[amssymb]{SIunits}
\usepackage{graphicx}
\usepackage{color}
\usepackage{ulem}
\usepackage{enumerate}
\usepackage{bigdelim}
\usepackage{array}
\usepackage{tabularx}
\usepackage{multirow}

\PassOptionsToPackage{colorlinks=true,linkcolor=magenta,citecolor=green,urlcolor=blue}{hyperref}
\usepackage{hyperref}
\hypersetup{%
pdftitle={Quantification of finite-temperature effects on adsorption geometries of pi-conjugated molecules},
pdfauthor={Giuseppe Mercurio},
pdfsubject={Physical Review B},
pdfkeywords={adsorbate structure, X-ray standing waves, density functional theory, adsorbate vibrations}
}

\begin{document}
	
\preprint{%
published in Phys. Rev. B \textbf{88}, 035421 (2013)
DOI: \href{http://dx.doi.org/10.1103/PhysRevB.88.035421}{10.1103/PhysRevB.88.035421}
}

\title{Quantification of finite-temperature effects on adsorption geometries of $\pi$-conjugated molecules}

\author{G.~Mercurio}
\altaffiliation{\textit{Current address:} University of Hamburg and Center for Free-Electron Laser Science, Luruper Chausse 149, 22761 Hamburg, Germany; \href{mailto:giuseppe.mercurio@desy.de}{giuseppe.mercurio@desy.de}}
\affiliation{Peter Gr\"{u}nberg Institut (PGI-3), Forschungszentrum J\"{u}lich, 52425 J\"{u}lich, Germany}
\affiliation{J\"{u}lich Aachen Research Alliance (JARA), Fundamentals of Future Information Technology, 52425
J\"{u}lich, Germany}

\author{R.~J.~Maurer}
\affiliation{Department Chemie, Technische Universit\"{a}t M\"{u}nchen, Lichtenbergstr.~4, 85747 Garching,
Germany}

\author{W.~Liu}
\affiliation{Fritz-Haber-Institut der Max-Planck-Gesellschaft, Faradayweg 4-6, 14195 Berlin, Germany}

\author{S.~Hagen}
\affiliation{Freie Universit\"{a}t Berlin, Fachbereich Physik, Arnimallee 14, 14195 Berlin, Germany}

\author{F.~Leyssner}
\affiliation{Freie Universit\"{a}t Berlin, Fachbereich Physik, Arnimallee 14, 14195 Berlin, Germany}

\author{P.~Tegeder}
\affiliation{Freie Universit\"{a}t Berlin, Fachbereich Physik, Arnimallee 14, 14195 Berlin, Germany}
\affiliation{Physikalisch-Chemisches Institut, Ruprecht-Karls-Universit\"{a}t Heidelberg, Im Neuenheimer Feld
253, 69120 Heidelberg, Germany}

\author{J.~Meyer}
\affiliation{Department Chemie, Technische Universit\"{a}t M\"{u}nchen, Lichtenbergstr.~4, 85747 Garching,
Germany}

\author{A.~Tkatchenko}
\affiliation{Fritz-Haber-Institut der Max-Planck-Gesellschaft, Faradayweg 4-6, 14195 Berlin, Germany}

\author{S.~Soubatch}
\affiliation{Peter Gr\"{u}nberg Institut (PGI-3), Forschungszentrum J\"{u}lich, 52425 J\"{u}lich, Germany}
\affiliation{J\"{u}lich Aachen Research Alliance (JARA), Fundamentals of Future Information Technology, 52425
J\"{u}lich, Germany}

\author{K.~Reuter}
\affiliation{Department Chemie, Technische Universit\"{a}t M\"{u}nchen, Lichtenbergstr.~4, 85747 Garching,
Germany}

\author{F.~S.~Tautz}
\affiliation{Peter Gr\"{u}nberg Institut (PGI-3), Forschungszentrum J\"{u}lich, 52425 J\"{u}lich, Germany}
\affiliation{J\"{u}lich Aachen Research Alliance (JARA), Fundamentals of Future Information Technology, 52425
J\"{u}lich, Germany}

\date{\today}

\begin{abstract}
The adsorption structure of the molecular switch azobenzene on Ag(111) is investigated by a combination of 
normal incidence x-ray standing waves and dispersion-corrected density functional theory.~The inclusion of
non-local collective substrate response (screening) in the dispersion correction improves the description of
dense monolayers of azobenzene, which exhibit a substantial torsion of the molecule.~Nevertheless, for a
quantitative agreement with experiment explicit consideration of the effect of vibrational mode anharmonicity
on the adsorption geometry is crucial.
\end{abstract}


\maketitle

Precise experimentally determined structures of large organic adsorbates are indispensable --- for the
detailed understanding of their wide-ranged functionalities, but also for the benchmarking of ab initio
electronic structure calculations \cite{Tautz07PiSS82_479,Romaner07PRL99_256801,Koch08JotACS130_7300}.~For
large molecules with polarizable $\pi$-electron systems, van der Waals (vdW) interactions are substantial and
may critically influence the adsorption geometry
\cite{Atodiresei09PRL102_136809,Mittendorfer11PRB84_201401,Liu12TJoPCL3_582,Sony07PRL99_176401}.~Accounting
for these interactions in ab initio calculations remains a challenge, and different approaches to this problem
at varying degrees of accuracy are currently explored
\cite{Grimme10TJoCP132_154104,Lee10PRB82_81101,Klimevs10JoPCM22_22201,Steinmann11JoCTaC7_3567,
Tkatchenko09PRL102_73005,Tkatchenko10MB35_435,Ruiz12PRL108_146103}.~Due to the system sizes inherent
to large molecular adsorbates, efficient semi-empirical dispersion correction schemes to density-functional
theory (SEDC-DFT) are particularly promising \cite{Tkatchenko10MB35_435}.~However, their approximate nature
makes them even more dependent on reliable experimental benchmark structures.~This holds in particular for
adsorption at metal surfaces, where the non-local collective substrate response (many-body electronic
screening) requires advancements beyond the traditional pairwise summation of vdW interactions in these
schemes \cite{Ruiz12PRL108_146103,Mercurio10PRL104_36102}.

With SEDC-DFT now striving for the approximate inclusion of the collective substrate response
\cite{Ruiz12PRL108_146103}, the accuracy increases to approximately \unit{0.1}{\angstrom} for the predicted
adsorption heights \cite{Ruiz12PRL108_146103,Liu12PRB86_245405,Al-Saidi12NL12_997,Wagner12PRL109_76102}.~At
this level of accuracy, a new issue arises:~Experiments for structure determination are often carried out
close to room temperature, while in SEDC-DFT the ground state (at \unit{0}{\kelvin}) is normally
calculated.~The complex internal vibrational structure of large organic adsorbates which may sensitively
influence the experimental time-averaged geometry is thus neglected.~In this Letter we show that the inclusion
of such thermal expansion effects into SEDC-DFT is indeed necessary to reach quantitative agreement between
experiment and theory.~Hence, benchmarking at the current level of sophistication requires the careful
analysis of finite-temperature effects.~Otherwise misleading conclusions with respect to the SEDC-DFT accuracy
might be obtained.
\begin{figure}[!b]
\includegraphics[scale=0.69]{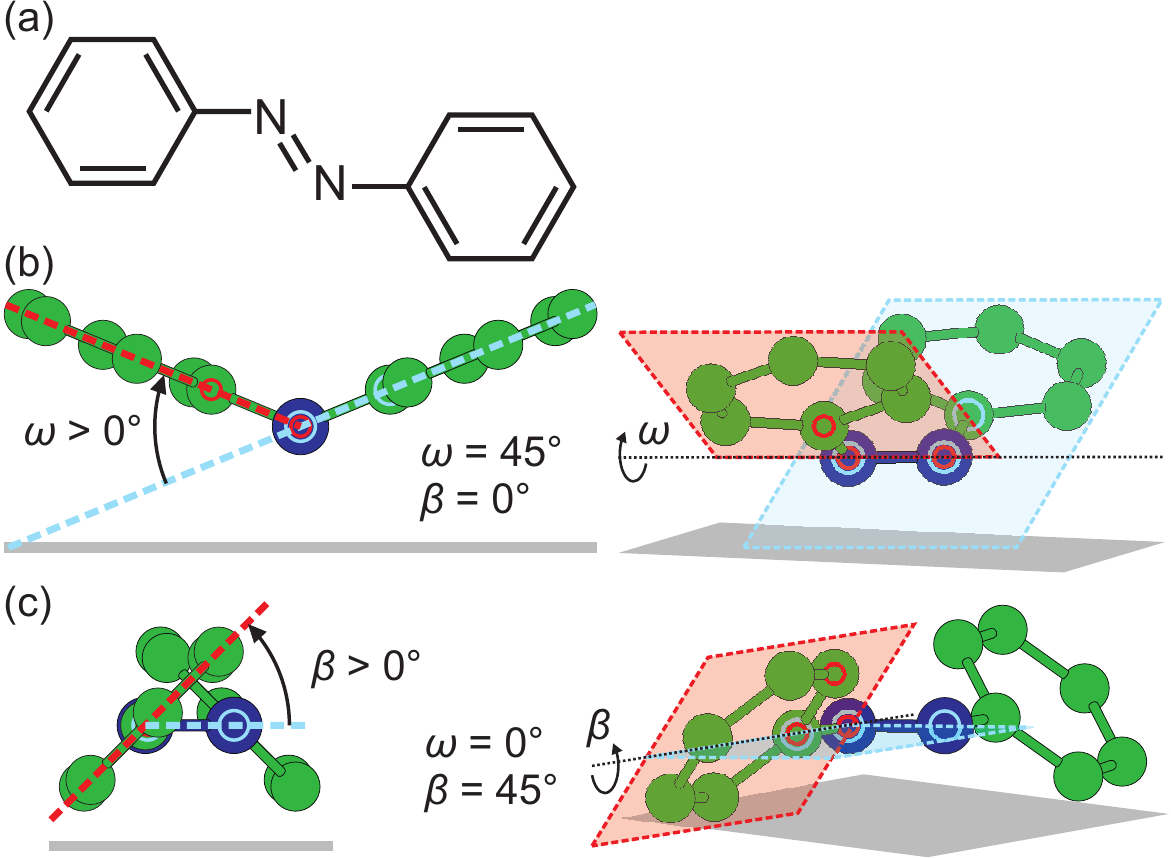}
\caption{(a)~Structure formula of azobenzene (AB).~(b)~Side view and perspective view of AB with $\omega
= 45\degree$ and $\beta = 0\degree$.~(c)~Side view and perspective view of AB with $\omega
= 0\degree$ and $\beta = 45\degree$.~$\omega$ and $\beta$ are defined as the dihedral angles CNNC and CCNN
\cite{footnote_dihedral}, respectively.~The planes containing atoms C,N,N (panel b) and C,C,N (panel c) are
marked in red and the corresponding atoms are indicated by red circles.~The planes containing atoms N,N,C
(panel b) and C,N,N (panel c) are marked in light blue and the corresponding atoms are indicated by light
blue circles.~C atoms:~green spheres.~N atoms:~blue spheres.~For clarity, H atoms are not drawn.}
\label{Fig-1:angles-w-b}
\end{figure}

Our experiments have been carried out on azobenzene (AB, cf.~Fig.~\ref{Fig-1:angles-w-b}a) adsorbed at
Ag(111), by the normal incidence x-ray standing wave technique (NIXSW).~AB is a widely-used molecular switch
\cite{Molecular_Switches_Feringa}.~Investigations of its substrate interaction are driven by the challenge to
preserve the switching functionality in the presence of a surface.~In this context, the knowledge of the
adsorption structure is essential.~NIXSW is an established method to determine the adsorption geometry (in
particular adsorption heights) of large organic adsorbates
\cite{Zegenhagen93SSR18_202,Woodruff05RoPiP68_743}.~The AB/Ag(111) system has been studied by NIXSW
before and the results were compared to the SEDC-DFT approaches of the time to conclude on the importance of
(then untreated) electronic screening effects \cite{Mercurio10PRL104_36102}.~Using a most recent SEDC-DFT
revision that approximately includes non-local collective substrate response we here confirm this
proposition.~Also, accounting in our refined analysis for coverage dependence, we nevertheless show that the
crucial missing link to achieve quantitative agreement with experiment lies not on the electronic structure
level, but in hitherto generally neglected finite-temperature effects.

NIXSW experiments were carried out at ESRF, beamline ID32, under ultra-high vacuum conditions
(\unit{\approx5}{\times\power{10}{-10}\milli\bbar}) \cite{Mercurio10PRL104_36102}.~The Ag(111) surface was
cleaned by several cycles of Ar$^{+}$ ion sputtering and annealing at \unit{820}{\kelvin}.~Multilayers of
AB were deposited from an effusion cell held at room temperature onto the atomically ordered Ag(111) crystal
at \unit{220}{\kelvin}.~AB monolayers were then prepared by desorption from multilayers, by heating with a
rate of \unit{1}{\kelvin\per\second} until the multilayer desorption peak had decayed and before the
monolayer peak was observed \cite{Mercurio_PhD_Thesis_conference}.~This guarantees coverages close to one
monolayer.~The AB fragment mass of \unit{77}{amu} (C$_6$H$_{5}^{+}$) was monitored on-line with a quadrupole
mass spectrometer.~The Ag crystal was kept at \unit{210}{\kelvin} during the NIXSW experiments to prevent
desorption.

Vibrations are expected to influence the average geometry of the adsorbate (via vibrational mode
anharmonicity) and to broaden the distribution of atoms around their average positions (via vibrational
dynamics) \cite{Zegenhagen93SSR18_202}.~While this will affect both the coherent position $P_c$ and the
coherent fraction $F_c$ of the NIXSW signal, prevalent (harmonic) Debye-Waller theory only considers
temperature effects on $F_c$ \cite{Zegenhagen93SSR18_202,Woodruff94JoPCM6_10633}.~$P_c$ defines the average
adsorption height of a species, while $F_c$ quantifies the corresponding height
distribution.~A coherent fraction of 1 means that all photoemitters of a certain species have precisely the
same adsorption height above the relevant family of Bragg planes, while a coherent fraction of 0 indicates a
homogeneous distribution of the photoemitters throughout the Bragg spacing.~In general, $F_c < 1$ due to
unavoidable structural disorder \cite{Mercurio13PRB87_45421}, adsorbate and substrate thermal
vibrations \cite{Supplementary}.~However, the coherent fractions of different
chemical species also contain information about the internal geometry of the adsorbate that has so far been
left aside in most NIXSW studies.~Here we recover this information by including differences between the $F_c$
of different species into our analysis.
\begin{figure}[!b]
\includegraphics[scale=0.75]{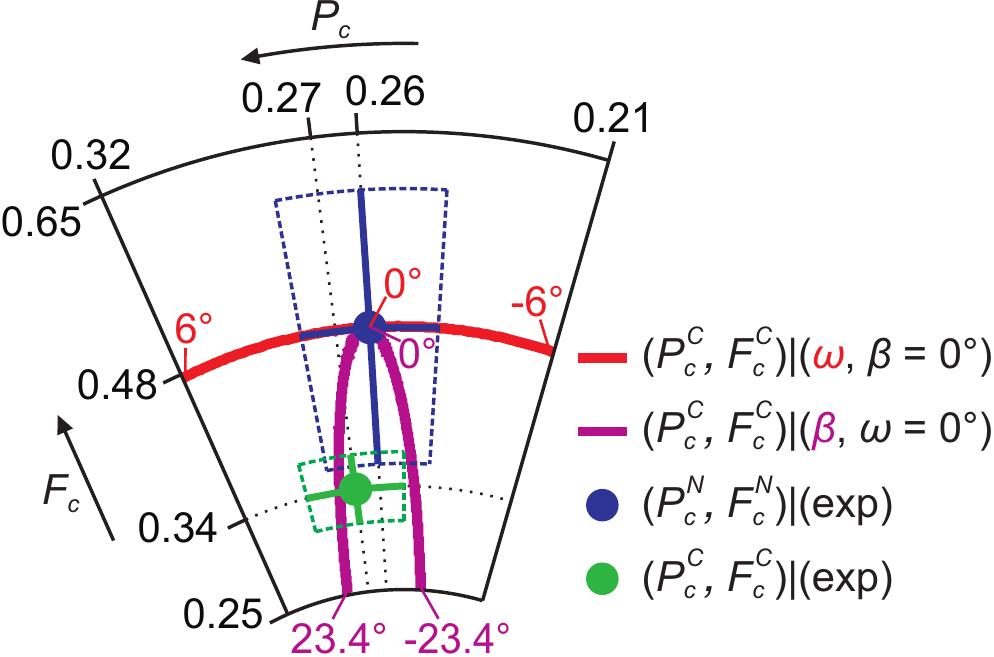}
\caption{Argand diagram indicating the NIXSW experimental results and NIXSW simulations for
AB/Ag(111).~($\bullet$, green):~Average experimental ($P_c$, $F_c$)=($0.27\pm0.02$, $0.34\pm0.03$) of
carbon.~($\bullet$, blue):~Average experimental ($P_c$, $F_c$)= ($0.26\pm0.02$, $0.48\pm0.12$) of
nitrogen.~Green and blue solid lines:~Corresponding error bars.~Green and blue dashed lines:~Corresponding
error regions \cite{Supplementary}.~Red solid line:~Simulated ($P_c$, $F_c$) of carbon with $-6\degree \leq
\omega \leq 6\degree$ and $\beta = 0\degree$.~Magenta solid line:~Simulated ($P_c$, $F_c$) of carbon with
$-23.4\degree \leq \beta \leq 23.4\degree$ and $\omega = 0\degree$.}
\label{Fig-2:Argand-diagram}
\end{figure} 

In the present case of AB/Ag(111), NIXSW provides coherent positions $P_{c}^{\mathrm{C}}=0.27 \pm
0.02$, $P_{c}^{\mathrm{N}}= 0.26 \pm 0.02$ and coherent fractions $F_{c}^{\mathrm{C}}= 0.34 \pm
0.03$, $F_{c}^{\mathrm{N}}=0.48 \pm 0.12$ (Fig.~\ref{Fig-2:Argand-diagram})
\cite{footnote_differences}.~Therefore, while the respective coherent positions are identical within the
errors, the coherent fraction of C is $29 \%$ smaller than the one of N.~In our
refined structure determination, we ascribe this difference to the internal geometry of AB, assuming that
$F_{c}^{\mathrm{C}_i}$ and $F_{c}^{\mathrm{N}_i}$, the coherent fractions of individual C and N atoms, are
equal (and smaller than 1 due to disorder) \cite{Supplementary}.~To solve the AB structure, two internal
degrees of freedom are considered:~the tilt angle $\omega$ (Fig.~\ref{Fig-1:angles-w-b}a,
\cite{footnote_omega}) and the torsion angle $\beta$ (Fig.~\ref{Fig-1:angles-w-b}b), defined as dihedral
angles CNNC and CCNN \cite{footnote_dihedral}, respectively.~It is impossible to explain the ratio
$F_{c}^{C}/F_{c}^{N}$ in a model in which $\omega$ is the only internal degree of freedom of the molecule,
because any distortion along $\omega$ that would lead to a decrease of $F_{c}^{\mathrm{C}}$ would at the same
time result in an increase of the coherent position $P_{c}^{\mathrm{C}}$ which is related to the average
adsorption height of the carbon atoms.~Hence, an additional degree of freedom must be considered to explain
the measured NIXSW structure parameters.~A plausible choice is the torsion angle $\beta$, because for small
angles $\omega$ a finite $\beta$ would broaden the carbon distribution essentially without changing the
average carbon height (Fig.~\ref{Fig-2:Argand-diagram} magenta curve).~Note that this broadening could in
principle be due to a \textit{static} distortion of the molecule and/or due to its vibrational
\textit{dynamics}.~However, for a purely dynamical reduction of the average coherent fraction
$F_{c}^{\mathrm{C}}$ by \unit{29}{\%} an unreasonably large C vibrational amplitude of the order
\unit{\pm0.30}{\angstrom} (with fixed N atoms) would be required.~Therefore, we will first consider a static
distortion before
coming back to a possible dynamical contribution.

Requiring $F_{c}^{\mathrm{C}_i}=F_{c}^{\mathrm{N}_i}$ and constructing the molecular geometry such that the
measured values for $P_{c}^{\mathrm{C}}$, $P_{c}^{\mathrm{N}}$, $F_{c}^{\mathrm{C}}$, $F_{c}^{\mathrm{N}}$ are
obtained, we find an adsorption geometry with $d_{\mathrm{N- Ag}}$ of \unit{2.97 \pm 0.05}{\angstrom}, a tilt
angle $\omega$ of \unit{-0.7}{\degree} and a torsion angle $\beta$ of \unit{17.7}{\degree} from our NIXSW data
(cf.~Table~\ref{Table:summary-results}) \cite{footnote_posnegbeta,derivation}.

\begin{figure}[!t]
\includegraphics[scale=0.85]{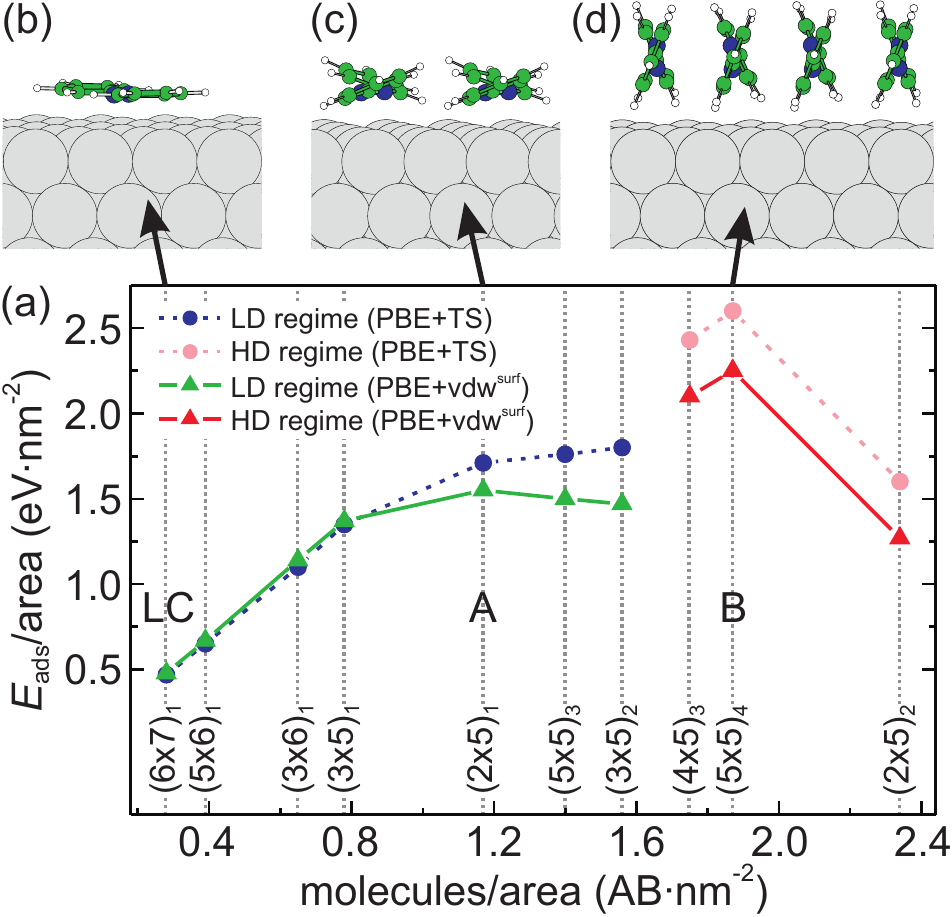}
\caption{(a)~Adsorption energy per surface area vs.~surface coverage of AB/Ag(111) as calculated with
PBE+vdW$^{\rm surf}$ ($\blacktriangle$, solid line) and PBE+TS ($\bullet$, dashed line).~The low density (LD)
and high density (HD) coverage regimes (see text) are indicated by different colors (green/blue and
red/pink).~Corresponding unit cells and the numbers $n$ of AB molecules in the unit cell are given as
(X$\times$Y)$_n$.~Adsorption geometry (vdW$^{\rm surf}$) of the low-coverage (LC) phase (b), of phase
A (c) and of phase B (d).}
\label{Fig-3:ads-energy}
\end{figure}

A torsion angle $\beta$ of more than \unit{17}{\degree} is difficult to rationalize for a single molecule
adsorbed on the surface without neighbors.~Yet, all calculations so far
\cite{Mercurio10PRL104_36102,Li12PRB85_121409,McNellis09PRB80_205414} have been carried out for
single molecules (while our experiment is performed on a condensed layer, see above).~We therefore need to
analyze the coverage- and packing-dependence of the adsorption geometry of AB/Ag(111) theoretically.~While our
previous SEDC-DFT calculations~\cite{Mercurio10PRL104_36102} for this system were carried out at the level of
the dispersion-correction scheme by Tkatchenko and Scheffler (TS) \cite{Tkatchenko09PRL102_73005}, we now
employ the more recent vdW$^\mathrm{surf}$ scheme \cite{Ruiz12PRL108_146103}, which accounts for non-local
collective substrate response via renormalization of the dispersion coefficients on the basis of
Lifshitz-Zaremba-Kohn theory.~Details of the calculations can be found in the supplement
\cite{Supplementary}.~We determine the optimized adsorption geometries for a range of different surface
unit-cells \cite{Supplementary}, with one AB per $(6 \times 7)$ cell representing the low-coverage (LC) limit
and two AB molecules in a $(2 \times 5)$ cell leading to the highest considered molecular surface density
(cf.~Fig.~\ref{Fig-3:ads-energy}).
 
The PBE+vdW$^{\rm surf}$ results compiled in Fig.~\ref{Fig-3:ads-energy} show that the adsorption geometry
indeed varies substantially with increasing molecular surface density.~While in the LC limit the adsorbed
molecule is essentially flat (Fig.~\ref{Fig-3:ads-energy}b), the tilt and torsion angles $\omega$ and $\beta$
increase with the packing density (Fig.~\ref{Fig-3:ads-energy}c-d).~As a consequence of the internal
distortion of the molecule, the vertical adsorption height of the azo-bridge also increases
\cite{Supplementary};~this tendency continues beyond the critical coverage of \unit{1.56}{AB\cdot nm^{-2}} at
which the now nearly upright molecules start to flatten out again within the increasingly dense overlayer
(Fig.~\ref{Fig-3:ads-energy}d).~The flattening of the upright molecule implies an increasingly asymmetric
position of the azo-bridge, i.e.~different vertical adsorption heights of the two nitrogen atoms, with a
consequent lifting from the surface \cite{Supplementary}.

With most of the adsorption energy of the flat AB molecule in the LC limit coming from dispersive interactions
with the substrate, the binding energy \textit{per molecule} naturally decreases in the distorted high density
structures (cf.~supplementary Table~I \cite{Supplementary}).~Due to the denser packing, the adsorption energy
\textit{per surface area} $E_{\mathrm{ads}}$/area
nevertheless increases and reaches a maximum at \unit{1.17}{AB\cdot nm^{-2}}
(Fi.g~\ref{Fig-3:ads-energy}a).~The intermolecular vdW
interactions in the high density phases further increase $E_\mathrm{{ads}}$/area, which reaches a second
maximum at a density of \unit{1.87}{AB\cdot nm^{-2}}.~Our calculations therefore predict the existence of two
optimum packing densities, a phase A (Fig.~\ref{Fig-3:ads-energy}c) at \unit{1.17}{AB\cdot nm^{-2}} and a
phase B (Fig.~\ref{Fig-3:ads-energy}d) at \unit{1.87}{AB\cdot nm^{-2}}.~Qualitatively similar findings are
obtained with the TS scheme.~In detail, however, there are decisive differences.~For example, TS fails to
predict the maximum of $E_{\mathrm{ads}}$/area corresponding to phase A, cf.~Fig.~\ref{Fig-3:ads-energy}a.

To decide which structure --- if any of the above --- we have in our experiment, we take the calculated ground
state geometries and compare them to experiment (Table~\ref{Table:summary-results}).~Phase LC can be ruled
out, both because of its small torsion angle, and because of our sample preparation procedure which yields
dense layers.~The average adsorption height of N atoms in phase B is \unit{4.39}{\angstrom}, which is
inconsistent with the experimental value of \unit{2.97}{\angstrom} or --- modulo a Bragg
spacing --- \unit{5.32}{\angstrom}.~We therefore conclude that our NIXSW experiment has been carried out on a
structure similar to phase A.~This conclusion is consistent with the expectation that neither of the two
dispersion-correction schemes will work reliably at the packing density of phase B that is close to the one of
the molecular crystal, in which even the vdW$^\mathrm{surf}$ scheme will overestimate lateral
interactions~\cite{Tkatchenko08PRB78_45116}, because higher-order many-body terms are
neglected~\cite{Tkatchenko12PRL108_236402}.~This neglect will contribute to a spurious stabilization of phase
B in the calculation.

In Table~\ref{Table:summary-results} the geometry parameters of phase A are summarized.~At \unit{0}{\kelvin}
the vdW$^\mathrm{surf}$ scheme yields a height of $d_{\mathrm{N- Ag}}=\unit{2.81}{\angstrom}$ at tilt
$\omega=11.7\degree$ and torsion $\beta=15.4\degree$, while the TS scheme predicts
$d_{\mathrm{N-Ag}}=\unit{3.26}{\angstrom}$, $\omega=7.5\degree$, and $\beta=18.6\degree$.~With regard to
$\beta$, we observe a good agreement of the ground state calculation with the experimental result
($\beta=17.7\degree$).~In contrast, the calculated adsorption heights of the azo-bridge are
\unit{0.16}{\angstrom} too small for vdW$^\mathrm{surf}$ and \unit{0.29}{\angstrom} too large for TS.~It is
clear that the inclusion of collective substrate response has a large impact on the predicted adsorption
height $d_{\mathrm{N-Ag}}$.~It tends to improve the TS prediction, although the height is still not
perfect, and the calculated $\omega$ is too large.

We will now show that the predictions of the vdW$^\mathrm{surf}$ theory can be substantially improved toward
a quantitative agreement with experiment, if the effect of finite temperature is taken into account.~In
particular, anharmonic contributions to the vibrational motion may modify the time-averaged geometries that
are experimentally observable \cite{Zegenhagen93SSR18_202}.~We demonstrate this by explicitly calculating the
harmonic vibrations for the adsorbed AB molecule at the optimum density of \unit{1.17}{AB\cdot nm^{-2}}, both
at PBE+vdW$^{\mathrm{surf}}$ and PBE+TS levels.~We then map out the anharmonic regimes of these modes at
energies around the experimentally employed \unit{210}{\kelvin}, by distorting the molecule along the
corresponding vibrational eigenvectors.~Next, we fit a Morse potential \cite{Slater57N180_1352} to these data
points for every harmonic mode and integrate the motion in the Morse potentials analytically to obtain the
shifts of the average positions at \unit{210}{\kelvin} relative to the harmonic minima.~Summing these shifts
over all vibrational modes, we finally construct an anharmonically corrected geometry for the adsorbed AB
molecule \cite{Supplementary}.

With this procedure we arrive at the following finite-temperature (\unit{210} {\kelvin}) structures
for the vdW$^\mathrm{surf}$ (TS) schemes:~$d_{\mathrm{N-Ag}}=\unit{2.98}{\angstrom}\,(\unit{3.23}
{\angstrom})$, $\omega=\unit{9.0}{\degree}\,(\unit{8.8}{\degree})$ and $\beta =
\unit{17.7}{\degree}\,(\unit{17.3}{\degree})$ (Table~\ref{Table:summary-results}).~Driven particularly by the
low-energy adsorbate-substrate stretching modes, anharmonic effects primarily affect
$d_{\mathrm{N-Ag}}$.~In the case of vdW$^\mathrm{surf}$, they lift the azo-bridge by \unit{0.17} {\angstrom}
into almost perfect agreement with the measured value of \unit{2.97 \pm 0.05} {\angstrom}.~At the same time,
the larger vertical adsorption height of the azo-bridge allows the molecule to flatten out again under the
influence of the van der Waals interaction with the metal (reduction of $\omega$), and to twist further as a
result of intermolecular interactions (increase of $\beta$).~Both tendencies bring the calculated geometry
closer to experiment, although the calculated $\omega$ remains too large.~For TS, on the other hand,
anharmonicity affects $d_{\mathrm{N-Ag}}$ and $\beta$ only mildly, because $d_{\mathrm{N-Ag}}$ is too large
even in the \unit{0}{\kelvin} calculation;~moreover, it has an adverse effect on $\omega$, because it brings
the molecule closer to the surface.~Overall, the quality gap between vdW$^{\mathrm{surf}}$ and TS is therefore
widened by the inclusion of anharmonic effects.

To check the self-consistency of the finite-temperature geometry, we have simulated NIXSW results on its
basis, with the aim to evaluate the influence of vibrational excitations on the coherence of the NIXSW signal
(cf.~supplementary Section III \cite{Supplementary} for details).~Note that our experimental values for
$\omega$ and $\beta$ in Table~\ref{Table:summary-results} are based on the assumption that
$F_{c}^{\mathrm{C}}$ and $F_{c}^{\mathrm{N}}$ are different exclusively due to static distortions.~For the
anharmonically corrected average
structure of the molecule, the NIXSW simulation yields a
$F_{c}^{\mathrm{C}}/F_{c}^{\mathrm{N}}=0.60\,(0.61)$~(TS in brackets), a value very close to both experiment
($0.71$) and the \unit{0}{\kelvin} structure (0.61 in vdW$^{\rm surf}$).~Most importantly, the reduction of
the coherent fractions due to vibrational motion around the average structure is similar for C and N, and
approximately equal to $10\%$ ($5\%$)~(TS in brackets), such that $F_{c}^{\mathrm{C}}/F_{c}^{\mathrm{N}}$
becomes $0.63$ ($0.62$) (cf.~supplementary Table~IV
\cite{Supplementary}), hence closer to experiment.~The nearly equal reduction of $F_{c}^{\mathrm{C}}$
and $F_{c}^{\mathrm{N}}$ due to thermal vibrations confirms \textit{a posteriori} that in deriving the
experimental structure we can
interpret the different experimental $F_{c}$s as being due to static distortion, and hence the experimental
geometry
parameters in Table~\ref{Table:summary-results} are the correct reference point for the calculated
finite-temperature geometry.
\begin{table}[!t]
\caption{Summary of the geometry parameters $d_{\mathrm{N-Ag}}$, $\omega$, and $\beta$ calculated by the two
different SEDC-DFT schemes PBE+TS and PBE+vdW$^{\text{surf}}$ for phase LC at \unit{0}{\kelvin},
and phase A at \unit{0}{\kelvin} and \unit{210}{\kelvin}.~Also shown are the experimental NIXSW
results.~Details concerning the calculation of $\beta$ and of all experimental error bars are reported in the
supplementary material \cite{Supplementary}.}
\begin{tabular}{c|cccc}
\hline\hline 
 & & $d_{\mathrm{N-Ag}}$ ({\AA})& $\omega$ ($^{\circ}$) &
$\beta$ ($^{\circ}$)
\tabularnewline
\hline
phase LC & TS  & $2.95$ & $1.8$ & $-0.6$ \tabularnewline
($T=\unit{0}{\kelvin}$) & vdW$^{\mathrm{surf}}$ & $2.61$ & $4.5$ & $-2.0$ \tabularnewline
\hline
phase A & TS  & $3.26$ & $7.5$ & $18.6$ \tabularnewline
($T=\unit{0}{\kelvin}$) & vdW$^{\mathrm{surf}}$   & $2.81$ & $11.7$ & $15.4$ \tabularnewline
\hline
phase A & TS  & $3.23$ & $8.8$ & $17.3$ \tabularnewline
($T=\unit{210}{\kelvin}$) & vdW$^{\mathrm{surf}}$   & $2.98$ & $9.0$ & $17.7$ \tabularnewline
 \hline
experiment & \multirow{2}{*}{NIXSW} & $2.97$ & $-0.7$ & $17.7$ \tabularnewline
($T=\unit{210}{\kelvin}$) & & $\pm 0.05$ & ($+2.3$/$-2.2$) & ($+2.4$/$-2.7$) \tabularnewline
\hline\hline
\end{tabular}
\label{Table:summary-results}
\end{table}

In conclusion, we have analyzed the structure of the archetypal molecular switch azobenzene on the Ag(111)
surface.~We find that the inclusion of collective substrate response into SEDC-DFT correction schemes is
absolutely essential for a correct description of the adsorption geometry.~However, since the
vdW$^{\mathrm{surf}}$ scheme leads to a reduction of both the dispersion coefficients \textit{and} the van der
Waals radii, it may --- counterintuitively --- decrease the adsorption height compared to SEDC-DFT without
collective substrate response.~This is clearly observed for the LC phase.~However, we have identified two
effects which increase the adsorption height again.~First, this is the dense molecular packing and the
associated molecular distortion, which increase $d_{\mathrm{N-Ag}}$ by \unit{0.20}{\angstrom}.~Second, the
anharmonicity of molecular vibrations raises $d_{\mathrm{N-Ag}}$ by another \unit{0.17}{\angstrom}.~The
remaining disagreements between experiment and theory notwithstanding, there are three important findings
which can be generalized:~Firstly, information regarding the molecular conformation beyond the average
positions
of certain chemical species can be retrieved from the coherence of the NIXSW signal with suitable
simulations.~Secondly, thermal expansion due to the anharmonicity of molecular vibrations not captured in
Debye-Waller theory \cite{Zegenhagen93SSR18_202} may contribute \unit{0.1-0.2}{\angstrom} to the adsorption
height.~This must be taken into account in future benchmarks of high-level ab initio theory against NIXSW,
either by carrying out the experiments at low temperature or by inclusion of finite-temperature vibrational
effects into the calculation, as sketched in the present paper.~And thirdly, our observation that all three
geometry parameters $d_{\mathrm{N-Ag}}$, $\omega$, and $\beta$ develop into the correct direction if
anharmonic effects are included proves that the PBE+vdW$^{\mathrm{surf}}$ scheme captures the essential
physics of both chemical \textit{and} dispersion interactions and is therefore a good starting point as a
ground state calculation for the adsorption of large $\pi$-conjugated molecules.

We acknowledge financial support of the Deutsche Forschungsgemeinschft TA244/3-2, SFB 658 and RE1509/16-1.


\begin{thebibliography}{36}
\bibitem [{foo({\natexlab{c}})}]{footnote_omega}%
  \BibitemOpen
  \bibinfo {note} {Note that in Ref.~\cite{Mercurio10PRL104_36102} we used a
  different tilt angle $\tilde{\omega}$ defined as the the angle between the
  N-C bond of AB and the surface plane.}\BibitemShut {Stop}%
\end{thebibliography}
%

\end{document}